\documentclass[12pt,twoside]{report} 

\usepackage{setspace} 
\usepackage[utf8]{inputenc}
\usepackage[english]{babel} 
\usepackage{wrapfig}
\usepackage{graphicx} 
\usepackage{float}
\usepackage{arabtex} 
\usepackage[version=4]{mhchem} 
\usepackage{amsmath} 
\usepackage{amssymb} 
\usepackage{textcomp,gensymb} 
\usepackage{colortbl}
\usepackage{rotating} 
\usepackage{bm} 
\usepackage{multirow}
\usepackage{longtable} 
\usepackage{booktabs} 
\usepackage{acronym}
\usepackage{pdfpages} 
\usepackage{caption}
\usepackage[nottoc]{tocbibind} 
\usepackage{subcaption} 
\PassOptionsToPackage{hyphens}{url}\usepackage{hyperref}
\usepackage{hyperref}
\usepackage{listings}
\usepackage[table]{xcolor} 
\definecolor{Gray}{gray}{0.9} 
\usepackage{listings}    
\usepackage{courier}

\definecolor{arduinoGreen}    {rgb} {0.17, 0.43, 0.01}
\definecolor{arduinoGrey}     {rgb} {0.47, 0.47, 0.33}
\definecolor{arduinoOrange}   {rgb} {0.8 , 0.4 , 0   }
\definecolor{arduinoBlue}     {rgb} {0.01, 0.61, 0.98}
\definecolor{arduinoDarkBlue} {rgb} {0.0 , 0.2 , 0.5 }

\lstdefinelanguage{Arduino}{
  language=C++, % begin with default C++ settings 
  keywordstyle=\color{arduinoGreen},   
  deletekeywords={  % remove all arduino keywords that might be in c++
                break, case, override, final, continue, default, do, else, for, 
                if, return, goto, switch, throw, try, while, setup, loop, export, 
                not, or, and, xor, include, define, elif, else, error, if, ifdef, 
                ifndef, pragma, warning,
                HIGH, LOW, INPUT, INPUT_PULLUP, OUTPUT, DEC, BIN, HEX, OCT, PI, 
                HALF_PI, TWO_PI, LSBFIRST, MSBFIRST, CHANGE, FALLING, RISING, 
                DEFAULT, EXTERNAL, INTERNAL, INTERNAL1V1, INTERNAL2V56, LED_BUILTIN, 
                LED_BUILTIN_RX, LED_BUILTIN_TX, DIGITAL_MESSAGE, FIRMATA_STRING, 
                ANALOG_MESSAGE, REPORT_DIGITAL, REPORT_ANALOG, SET_PIN_MODE, 
                SYSTEM_RESET, SYSEX_START, auto, int8_t, int16_t, int32_t, int64_t, 
                uint8_t, uint16_t, uint32_t, uint64_t, char16_t, char32_t, operator, 
                enum, delete, bool, boolean, byte, char, const, false, float, double, 
                null, NULL, int, long, new, private, protected, public, short, 
                signed, static, volatile, String, void, true, unsigned, word, array, 
                sizeof, dynamic_cast, typedef, const_cast, struct, static_cast, union, 
                friend, extern, class, reinterpret_cast, register, explicit, inline, 
                _Bool, complex, _Complex, _Imaginary, atomic_bool, atomic_char, 
                atomic_schar, atomic_uchar, atomic_short, atomic_ushort, atomic_int, 
                atomic_uint, atomic_long, atomic_ulong, atomic_llong, atomic_ullong, 
                virtual, PROGMEM,
                Serial, Serial1, Serial2, Serial3, SerialUSB, Keyboard, Mouse,
                abs, acos, asin, atan, atan2, ceil, constrain, cos, degrees, exp, 
                floor, log, map, max, min, radians, random, randomSeed, round, sin, 
                sq, sqrt, tan, pow, bitRead, bitWrite, bitSet, bitClear, bit, 
                highByte, lowByte, analogReference, analogRead, 
                analogReadResolution, analogWrite, analogWriteResolution, 
                attachInterrupt, detachInterrupt, digitalPinToInterrupt, delay, 
                delayMicroseconds, digitalWrite, digitalRead, interrupts, millis, 
                micros, noInterrupts, noTone, pinMode, pulseIn, pulseInLong, shiftIn, 
                shiftOut, tone, yield, Stream, begin, end, peek, read, print, 
                println, available, availableForWrite, flush, setTimeout, find, 
                findUntil, parseInt, parseFloat, readBytes, readBytesUntil, readString, 
                readStringUntil, trim, toUpperCase, toLowerCase, charAt, compareTo, 
                concat, endsWith, startsWith, equals, equalsIgnoreCase, getBytes, 
                indexOf, lastIndexOf, length, replace, setCharAt, substring, 
                toCharArray, toInt, press, release, releaseAll, accept, click, move, 
                isPressed, isAlphaNumeric, isAlpha, isAscii, isWhitespace, isControl, 
                isDigit, isGraph, isLowerCase, isPrintable, isPunct, isSpace, 
                isUpperCase, isHexadecimalDigit, 
                }, 
  morekeywords={   % add arduino structures to group 1
                break, case, override, final, continue, default, do, else, for, 
                if, return, goto, switch, throw, try, while, setup, loop, export, 
                not, or, and, xor, include, define, elif, else, error, if, ifdef, 
                ifndef, pragma, warning,
                }, 
  keywordstyle=[2]\color{arduinoBlue},   
  keywords=[2]{   % add variables and dataTypes as 2nd group  
                HIGH, LOW, INPUT, INPUT_PULLUP, OUTPUT, DEC, BIN, HEX, OCT, PI, 
                HALF_PI, TWO_PI, LSBFIRST, MSBFIRST, CHANGE, FALLING, RISING, 
                DEFAULT, EXTERNAL, INTERNAL, INTERNAL1V1, INTERNAL2V56, LED_BUILTIN, 
                LED_BUILTIN_RX, LED_BUILTIN_TX, DIGITAL_MESSAGE, FIRMATA_STRING, 
                ANALOG_MESSAGE, REPORT_DIGITAL, REPORT_ANALOG, SET_PIN_MODE, 
                SYSTEM_RESET, SYSEX_START, auto, int8_t, int16_t, int32_t, int64_t, 
                uint8_t, uint16_t, uint32_t, uint64_t, char16_t, char32_t, operator, 
                enum, delete, bool, boolean, byte, char, const, false, float, double, 
                null, NULL, int, long, new, private, protected, public, short, 
                signed, static, volatile, String, void, true, unsigned, word, array, 
                sizeof, dynamic_cast, typedef, const_cast, struct, static_cast, union, 
                friend, extern, class, reinterpret_cast, register, explicit, inline, 
                _Bool, complex, _Complex, _Imaginary, atomic_bool, atomic_char, 
                atomic_schar, atomic_uchar, atomic_short, atomic_ushort, atomic_int, 
                atomic_uint, atomic_long, atomic_ulong, atomic_llong, atomic_ullong, 
                virtual, PROGMEM,
                },  
  keywordstyle=[3]\bfseries\color{arduinoOrange},
  keywords=[3]{  % add built-in functions as a 3rd group
                Serial, Serial1, Serial2, Serial3, SerialUSB, Keyboard, Mouse,
                },      
  keywordstyle=[4]\color{arduinoOrange},
  keywords=[4]{  % add more built-in functions as a 4th group
                abs, acos, asin, atan, atan2, ceil, constrain, cos, degrees, exp, 
                floor, log, map, max, min, radians, random, randomSeed, round, sin, 
                sq, sqrt, tan, pow, bitRead, bitWrite, bitSet, bitClear, bit, 
                highByte, lowByte, analogReference, analogRead, 
                analogReadResolution, analogWrite, analogWriteResolution, 
                attachInterrupt, detachInterrupt, digitalPinToInterrupt, delay, 
                delayMicroseconds, digitalWrite, digitalRead, interrupts, millis, 
                micros, noInterrupts, noTone, pinMode, pulseIn, pulseInLong, shiftIn, 
                shiftOut, tone, yield, Stream, begin, end, peek, read, print, 
                println, available, availableForWrite, flush, setTimeout, find, 
                findUntil, parseInt, parseFloat, readBytes, readBytesUntil, readString, 
                readStringUntil, trim, toUpperCase, toLowerCase, charAt, compareTo, 
                concat, endsWith, startsWith, equals, equalsIgnoreCase, getBytes, 
                indexOf, lastIndexOf, length, replace, setCharAt, substring, 
                toCharArray, toInt, press, release, releaseAll, accept, click, move, 
                isPressed, isAlphaNumeric, isAlpha, isAscii, isWhitespace, isControl, 
                isDigit, isGraph, isLowerCase, isPrintable, isPunct, isSpace, 
                isUpperCase, isHexadecimalDigit, 
                },      
  stringstyle=\color{arduinoDarkBlue},    
  commentstyle=\color{arduinoGrey},    
   numbers=left,                    
  numbersep=5pt,                   
  numberstyle=\color{arduinoGrey},    
  breaklines=true,                    % wordwrapping
  tabsize=2,         
  basicstyle=\ttfamily  
} 
\hypersetup{colorlinks=false} 
\usepackage[letterpaper, top=1in, bottom = 1in, right = 1in]{geometry}

\usepackage[nonumberlist,acronym]{glossaries}
\makeglossaries
\begin{document} 

\begin{figure}
    \centering
    \includegraphics[scale=1.5]{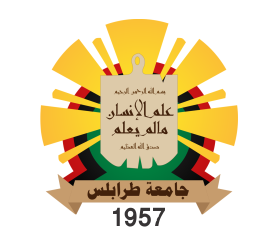}
\end{figure}

\begin{center}
{\Large \rm University of Tripoli \linebreak}
{\Large \rm Faculty of Engineering \linebreak}
{\Large \rm Electrical and Electronic Engineering Department \linebreak}
\vspace*{1cm}
\baselineskip 30pt
\end{center}
\vspace{-0.2cm}
\begin{center}
    {\Large \rm B.Sc. Project \linebreak}
\end{center} 
\vspace{-0.2cm}

\begin{center}
{\huge \rm \textbf{How Well Sensing Integrates with Communications in MmWave Wi-Fi?
%Evaluation of mmWave WiFi PHY Models for CSI Based WLAN Sensing  
}}% Sense-Fi:
\baselineskip 30pt
\end{center}
\begin{center}
\baselineskip 30pt
\center{\Large \rm Prepared by: Mohamed Hussein Abdalgader}\vspace*{-.5cm} \center{\Large \rm Supervised by: Dr. Nadia Adem}
\end{center}
\begin{center}
\baselineskip 30pt
{\large \rm Spring 2022 \linebreak}
{\large \rm Tripoli-Libya \linebreak}
\end{center}

\thispagestyle{empty}

\pagenumbering{roman} 

\chapter*{Acknowledgements} 
First I give my Thanks to “Allah”, Where if it wasn’t for his support and help, I wouldn’t have completed this project. 
\\I would like to extend my thanks and gratitude to my supervisor “Dr. Nadia Adem”, for her support, suggestions, patience and encouragement throughout the project.
\\I am deeply and forever obliged to my family for their  support and encouragement throughout my entire life.

\chapter*{Abstract}
%The development of integrated sensing and communication (ISAC) systems has recently gained interest for its capability to address a variety of issues, including the ability to share hardware, software, and spectrum resources as well as working to improve the co-existence for both sensing and communication systems. However, it can also provide new sensing applications for localization, tracking, and health care.
The development of integrated sensing and communication (ISAC) systems has recently gained interest for its ability to offer a variety of services including resources sharing 
and  new applications, for example,  localization, tracking, and health care related. 
While the sensing capabilities are offered through many technologies, rending to their wide deployments and the high frequency spectrum they provide and high range resolution, its accessibility through the  Wi-Fi networks IEEE 802.11ad and 802.11ay has been getting the interest of research and industry.
%While both the sub 7 GHz and 60 GHz frequency bands can be used for WLAN sensing, some of these applications are only available with the 60 GHz band due to its higher bandwidth and higher range resolution. Such systems include IEEE 802.11ad and 802.11ay. 802.11bf task group
 Even though there is  a dedicated standardization body, namely the 802.11bf task group,  working  on enhancing the Wi-Fi sensing performance, investigations are needed  to evaluate the effectiveness of  various  sensing techniques. 
 % simulation is required to evaluate the effectiveness of parameter estimation for various contributions and sensing techniques. 
  %
In this  project,  we, in addition to  surveying related literature, we evaluate the sensing performance of the millimeter wave (mmWave) Wi-Fi systems  by simulating a scenario of a human target using Matlab simulation tools. % by integrating a number of existing simulating models. 
In this analysis, we processed  channel estimation data using the short time Fourier transform (STFT). Furthermore, using a channel variation threshold method, we evaluated the performance while reducing feedback. 
%Our results show that using STFT with 32 window size produces better results than using FFT without windowing segmentation, and that increasing the window overlap gives better results, which we demonstrated using 50\% and 90\% overlap. We also demonstrated how threshold sensing affects performance by using 0.05 and 0.1 threshold levels, which reduce feedback measurements by 48\% and 77\%, respectively. We also demonstrated that beam-forming training sensing results are highly dependent on the AWV of the antenna; for this, we used 2x2, 2x8, and 8x8 antenna arrays.
%%%
Our findings indicate that using STFT window overlap can provide good tracking results, and that the reduction in  feedback measurements using 0.05 and 0.1 threshold levels reduces feedback measurements by 48\% and 77\%, respectively, without significantly degrading performance.
 %%%%%%%
 %The use of a common environment model (channel model) is required for comparing performance from various contributions, also The detected targets must be present in the environment. The purpose of this project is to evaluate the performance of 60 GHz WLAN sensing by simulating a scenario using tools for the physical layer and channel modeling. Our findings indicate that using STFT window overlap can provide good tracking results, and that reducing feedback measurement doesn't significantly degrade performance.

\tableofcontents 

\listoffigures 
%%%%%%%%%%%%%%%%%%%%%%%%%%%%%%%%%%%%%%%%%%%%%
\glsaddall
\newacronym{mmwave}{mmWave}{millimeter wave}
\newacronym{aoa}{AOA}{Angle of Arrival}
\newacronym{bft}{BFT}{Beamforming Training}
\newacronym{cir}{CIR}{Channel Impulse Response}
\newacronym{ddhc}{DDHC}{Data-Driven Hybrid Channel Model}
\newacronym{fov}{FOV}{Field of View}
\newacronym{agc}{AGC}{Automatic Gain Control}
\newacronym{awv}{AWV}{Antenna Weight Vector}
\newacronym{csi}{CSI}{Channel State Information}
\newacronym{dmg}{DMG}{Directional Multi-Gigabit}
\newacronym{edmg}{EDMG}{Enhanced Directional Multi-Gigabit}
\newacronym{kpi}{KPI}{Key Performance Indicator}
\newacronym{mac}{MAC}{Medium Access Control}
\newacronym{phy}{PHY}{Physical}
\newacronym{ppdu}{PPDU}{Physical Layer Protocol Data Unit}
\newacronym{rssi}{RSSI}{Received Signal Strength Indicator}
\newacronym{trn}{TRN}{Training Field}
\newacronym{trrs}{TRRS}{Time-Reversal Resonating Strength}
\newacronym{wi-fi}{Wi-Fi}{Wireless Fidelity}
\newacronym{wlan}{WLAN}{Wireless Local Area Network}
\newacronym{mcs}{MCS}{Modulation and Coding Scheme}
\newacronym{sc}{SC}{Single Carrier}
\newacronym{sta}{STA}{Station}
\newacronym{stft}{STFT}{Short-Time-Fourier-Transform}
\newacronym{pri}{PRI}{pulse repetition interval}
\newacronym{prf}{PRF}{pulse repetition frequency}
\newacronym{cpi}{CPI}{coherent processing interval}

\printglossary[type=\acronymtype]
\printglossary

%%%%%%%%%%%%%%%%%%%%%%%%%%%%%%%%%%%%%%%%%%%%

\chapter{Introduction}
\pagenumbering{arabic} 
Future wireless networks are expected to provide services for sensing. Communication and sensing systems are usually developed independent of one another and use different frequency bands. However, communication signals in future wireless systems tend to have high resolution in both the time and angle domains due to the use of millimeter wave and MIMO technologies, this makes it possible to provide high accuracy sensing using communication signals. In order to increase spectrum efficiency and lower hardware costs, it is therefore preferable to jointly design the sensing and communication systems, so that they can use the same hardware and frequency range.
%Sharing hardware and spectrum is a strong motivator for integrated sensing and communication 
 In addition, the widespread deployment and availability of wireless communications devices, combined with recent technological advancements,  provide an opportunity to enable new sensing applications. Sensing over wireless local area networks (WLAN) is an example of such communication systems, WLAN sensing can provide new technologies for location tracking, target identification and recognition and environment mapping.
Human presence detection is one of various use cases that motivates the development of WLAN sensing devices since it is essential for smart home implementation to reduce energy waste and improve user experience. Human activity recognition is another use case of WLAN sensing since it can be used to assist in understanding human behaviors and intentions including fall detection, gesture recognition, and security. WLAN sensing can also be used for analyzing human health conditions, such as sleep quality and heartbeat estimation.
\\There are a few commercial devices that has been developed to use WLAN sensing, for example project Soli. developed products that use sensing to understand human motions at various scales from heartbeat to the movements of the human body, Also ORIGIN has created multiple AI engines for multiple Wi-Fi sensing use cases such as detection of falls, motion and presence.
\\Despite the existence of a few devices that use WLAN sensing, several significant issues with WLAN sensing remain unresolved, including unified frameworks, effects on communication and trade-off performance bounds, and the most effective data acquisition and signal processing techniques. This shows clearly the need for standardization and system evaluation to address the potential gap between the available technologies and the ideal solutions.
\section{Related Work}
In the efforts of addressing the  Wi-Fi sensing standardization issues, the IEEE 802.11 Wireless LAN Next-Generation Standing Committee (WNG SC) first discussed the development of a WLAN sensing project in 2019. The Project Authorization Request (PAR) were created in order to improve sensing capabilities through IEEE 802.11 devices. In this regard, IEEE 802.11bf Task Group (TGbf) also works to develop an amendment that defines modifications to the IEEE 802.11 medium access control (MAC) and physical layer (PHY) to improve WLAN sensing operation in the sub 7GHz and above 45 GHz frequency bands~\cite{du2022overview}.
\\The TGbf presents multiple requirements for various WLAN sensing use cases including range resolution. According to the task group, some use cases such as gesture recognition are only valid in the mm-Wave band, the sub 7GHz band can provide more accurate results using machine learning, but the mm-Wave band WLAN standards such as IEEE 802.11ad and 802.11ay are still required for use cases that require high resolution~\cite{rangeres}.  The IEEE 802.11ad standard,  introduced in 2012, was the first  amendment to use the mmWave band to enable multi-Gbps throughput. As mmWave signals experience much higher path loss than sub 7GHz signals, the PHY and MAC layers defined by IEEE 802.11ad differ significantly from those defined by the legacy IEEE 802.11 standards.
\\New applications such as wireless virtual reality (VR), vehicle-to-x connectivity, mobile offloading, and indoor and outdoor wireless backhaul require higher throughput and reliability as well as lower latency than what 802.11ad can provide. To handle this concern, the IEEE 802.11ay Task Group was formed, in 2015,   defining PHY and MAC amendments to the 802.11 standard that enable 100 Gb/s communications in the 60 GHz band in order to meet the requirements of such diverse new applications~\cite{ghasempour2017ieee}.
\\WLAN sensing is classified into two types based on the characteristics of either the wireless signal received signal strength indicator (RSSI) or the channel state information (CSI). The RSSI, which measures the received signal strength at the receiver, has been previously widely adopted in WLAN sensing based on fingerprint and geometric model-based methods~\cite{yang2013rssi}~\cite{yuan2013estimating}. %
%
%
%RSSI can also be used to determine the distance between the transmitter and receiver using a basic path loss model\cite{kumar2009distance}. 
RSSI-based techniques are typically simple and inexpensive to implement.
In contrast to RSSI, though,  CSI can provide fine wireless channel information at the physical layer, since it stores channel amplitude and phase information across multiple subcarriers, allowing it to differentiate between multi-path characteristics.  {CSI} is used in communication systems to syncronize time and frequency, and it has proven to provide good sensing and tracking performance~\cite{ma2019wifi}~\cite{hang2019wish}.
\\Doppler processing, which is used for radar sensing, may be employed with CSI sensing since it can offer information about frequency shift. Fast Fourier transform can be used to extract the Doppler frequency shift and use it to track the target~\cite{tahmoush2011time}.
\\To support the development of sensing methods and understand the performance of a combined wireless communication and sensing system, the availability of modeling tools and data sets is essential. 
National Institute of Standards and Technology (NIST) provides a software to simulate 802.11ay physical layer.%, the software includes the main characteristics of the IEEE 802.11ay PHY, 
Their suggested software can use the preamble from the IEEE 802.11ay packet to estimate the channel, frequency correction, and time synchronization of a received signal. It also uses sensing processing to provide sensing results. NIST also provided software to model the environment~\cite{blandino2022tools}. 
%\blue{we should not over praise and describe .. we can save this later..i.e. statement in red has more details than needed here right? we are handling related work here..}
\\Since most IEEE 802.11bf use cases are based on human sensing, it is important to accurately represent the human target for simulations and system evaluations. The Boulic global human walking model has been successfully used in the design of radar sensing systems~\cite{chen2019micro}.
\\In our project, we aim to use the tools mentioned above to investigate how signal processing affects the performance of sensing. %We will also test how well suggested techniques and directional sensing perform.
\section{Objectives}
the objective of this project is to evaluate the performance of two physical layer models,
%the 802.11ay \red{EDMG PPDU} \blue{have they been introduced?} for sensing.
The objectives can be summarized as follows:
\begin{itemize}
%\item We reviewed sensing related literature for number of technologies including WLAN and radar and set up a framework for data acquiring and signal processing for sensing.
%\blue{is this a legitimate? If so revise and include it ..}
\item Generating a deterministic channel using the ray-tracing method including a walking human target using the quasi-deterministic software~\cite{qd}, which is being integrated with the human target model~\cite{boulic1990global} to create an environment for sensing performance evaluation.
%\red{make more clear indicating the the existing software (cite it) is being integrated with the another human target model (cite it) to create an environment for sensing performance evaluation..}
\item Using 802.11ay PHY software~\cite{software}, we investigated  the performance of sensing using IEEE 802.11 packets for various sensing processing settings,  the threshold sensing, and also angle estimation of directional sensing using various antenna arrays.
% for both TRN-R and TRN-T packets.
\end{itemize}
\section{Outline}
The remaining chapters are organized as follows. Chapter 2 gives   WLAN sensing basics and related  IEEE802.11bf standardization efforts. In chapter 3,  WLAN physical layer models, to be used for sensing performance evaluation,  are introduced.  In chapter 4 the sensing framework  is provided.  Chapter 5 presents the simulation settings and results. We conclude the project  in chapter 6.

\chapter{Sensing in Wireless Local Area Networks}
Wi-Fi has progressed from a low-rate replacement for Ethernet to one of the most successful wireless technologies ever created. Wi-Fi is  widely used to provide Internet access in almost any public or private space. The unprecedented throughput demands of next-generation multimedia applications, combined with the exponential increase in Wi-Fi devices, will inevitably result in complex wireless networking challenges. Wi-Fi devices are becoming more diverse, as well as more bandwidth demanding, ranging from personal computers to smartphones, televisions, tablets, and a variety of other devices. This extremely dense and heterogeneous concentration of WiFi devices
%, placed in almost every corner of our indoor environments, 
 will provide the ideal opportunity to continuously map the surrounding environment using Wi-Fi signals as sounding waveforms.
\\WLAN sensing  uses Wi-Fi signals to perform sensing tasks by leveraging prevalent Wi-Fi infrastructures and ubiquitous  signals. During their propagation, Wi-Fi radio waves in particular  penetrate and bend on the surface of objects. 
With proper signal processing, received Wi-Fi signals can be used to sense the environment, detect obstacles, and interpret target movement.
\section{Sensing  Technologies}
WLAN sensing differs greatly from existing sensing technologies such as visible light, ultrasound, radio frequency identification (RFID), and ultra wide band (UWB). 
\\The location of light sensors is estimated using visible lights, for example,  transmitted from light-emitting diode (LED) transmitters at known locations, which is highly dependent on the line of sight (LOS) channel between transmitters and receivers~\cite{keskin2018localization}. WLAN Sensing on the other hand, can provide better coverage due to the ability of radio signals to pass through walls and provide additional non-LOS (NLOS) information.
\\ visible light requires the use of specialized infrastructure, such as photo detectors and imaging sensors, which may result in high system costs~\cite{zhang2016towards}. WLAN sensing, on the other hand, allows for the reuse of existing Wi-Fi devices at a significantly lower cost. Following that, ultrasound based sensing employs an ultrasonic transceiver to record the time of flight (ToF) between the transmitter and receiver before determining their separation distance based on the speed of sound~\cite{luo2018ultrasonic}. However, unlike Wi-Fi signals, which are not harmful to health, ultrasonic signals can be harmful to infants and pets who are sensitive to high frequency sounds~\cite{lin2019rebooting}. Furthermore, the speed of sound varies significantly with humidity and temperature, making it less stable than WLAN sensing based on electromagnetic waves.
\\RFID requires the deployment of a separate infrastructure~\cite{motroni2021survey}, whereas Wi-Fi can leverage previously deployed network access infrastructure, allowing for large scale commercial use. Besides that, RFID systems must deploy dedicated arrays of passive RFID tags at targets, raising deployment costs~\cite{han2015twins}.
\\Finally, UWB positioning technologies transmit extremely short pulses over a relatively large bandwidth, larger than 500MHz, to track objects passively~\cite{alarifi2016ultra}, making them resistant to multi-path issues and allowing for accurate ToF estimation. However, the lack of large scale infrastructure deployment limits the use of UWB for sensing.
%An Overview on IEEE 802.11bf: WLAN Sensing.
\section{IEEE 802.11bf Standard}
Since existing WLAN sensing devices still face challenges. The development of new WLAN standards to better support WLAN sensing is necessary. Traditional WLAN networks were originally designed for data transmission rather than sensing. Related  performance metrics for example  throughput and latency  are not applicable to evaluate WLAN sensing performance. Furthermore, received wireless signals may exhibit amplitude and phase distortions, which can be partially compensated for in conventional communication standards through  equalization. The accuracy of this phase error compensation, however, is coarse and may not meet the sensing requirements. Thus, improved transmission protocols and sounding processes are required to improve WLAN sensing performance.
\\CSI in WiFi communication describes the channel properties of a communication link by reflecting how wireless signals propagate in a physical environment after diffraction, reflections, and scattering. CSI describes the phase shift of multi-path and amplitude attenuation on each subcarrier for a pair of transmitter and receiver antennas. Although IEEE 802.11 compliant CSI extraction methods exist, CSI is not always available from the user side in commercial WiFi devices. This is due to the fact that most WiFi chipset manufacturers keep CSI access as a private feature and do not provide consumers with a dedicated related function interface. Only a few commercial WiFi devices can access CSI data, but with very limited flexibility~\cite{csitool1}~\cite{csitool2}. This has undoubtedly impacted WLAN sensing research and development. As a result, a unified and flexible functional interface for channel measurements is required.
\\To address the aforementioned issues, it is emerging
to have new WLAN sensing standards to support sensing functions on WiFi devices in a timely and efficient manner while not significantly affecting communication performance. The amendment is expected to enable backward compatibility and coexistence with  legacy IEEE 802.11 devices operating in the same band. 
\subsection{802.11bf Key Performance Indicators (KPI)}
Unlike IEEE 802.11 standards, IEEE 802.11bf requires a new set of KPIs for WLAN sensing use cases such as presence detection, activity recognition, and human target localization and tracking. The set of  IEEE 802.11bf task group    KPIs for WLAN sensing is provided below~\cite{Usagemodel} along with their definitions. 
\begin{itemize}
 \item \textbf{Range coverage } defined as the maximum allowable distance from
a sensing station (STA) to the target, within which the signal to noise ratio (SNR) is above a pre-defined threshold such that the targets can be successfully detected.
 \item \textbf{Field of view (FOV)}, used as a metric that indicates a sensing device's angular coverage. In other words, it identifies the angle through which the STA performs sensing.
 \item \textbf{Range resolution} which is the minimum distance between two targets that a sensing STA can distinguish on the same direction but at different ranges.
 \item \textbf{Angular resolution (azimuth/elevation)} known as the minimum angle that the sensing STA can distinguish between two targets at the same range.
 \item \textbf{Velocity resolution} defined as the minimal velocity difference
between two objects that a sensing STA can distinguish.
 \item \textbf{Accuracy} which indicates the difference between the estimated range, angle, velocity of an object and the ground truth.
 \item \textbf{Probability of detection} which is the ratio of the number of correct predictions to the number of all possible predictions.
 \item \textbf{Latency} which presents the expected time taken to finish the sensing process.
 \item \textbf{Refresh rate} which is the frequency during which the sensing refresh takes place.
 \item \textbf{Number of simultaneous targets} referring to the number of
targets that can be detected simultaneously within the sensing area.
\end{itemize}
\subsection{IEEE 802.11bf Sensing Procedure}
\begin{figure}[h]
    \centering
    \includegraphics[scale =1]{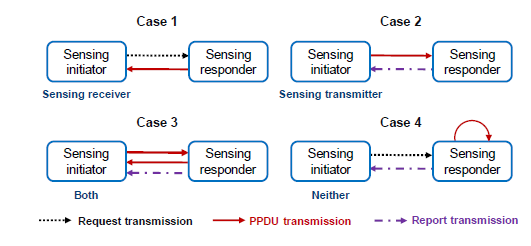}
    \caption{802.11bf sensing configuration cases~\cite{du2022overview}.}
    \label{case}
\end{figure}

The most important issue to be addressed within IEEE 802.11bf in terms of standardization is measurement acquisition, with the goal of obtaining sensing measurements from an IEEE 802.11 based transmission.
\\IEEE 802.11bf defines two terminologies: sensing procedure and sensing session~\cite{procedure}. The sensing procedure allows a STA to perform WLAN sensing and obtain measurement results, while the sensing session is an agreement between a sensing initiator and a sensing responder to participate in the sensing procedure. 
The sensing initiator and sensing responder are defined here based on which STA initiates a WLAN sensing procedure and requests and/or obtains measurements. 
A sensing initiator is a STA that initiates a sensing procedure, while a sensing responder is a STA that participates in a sensing procedure initiated by a sensing initiator. 
Both the sensing initiator and the sensing responder can be an access point (AP) or a non (AP) STA (i.e., client).
\\However, depending on who transmits the IEEE 802.11-based signal i.e., PPDU, which used to obtain measurements, two additional types of roles are available: sensing transmitter and sensing receiver. A sensing transmitter is a STA that sends PPDUs used for sensing measurements in a sensing procedure, while a sensing receiver receives PPDUs sent by a sensing transmitter and performs sensing measurements in a sensing procedure.
\\The ability to define each STA's role as a sensing transmitter or sensing receiver is a key feature of WLAN sensing. During a sensing procedure, a sensing initiator can be either a sensing transmitter or a sensing receiver, or both. Similar to the sensing initiator, the sensing responder can be either the sensing transmitter, the sensing receiver, or both. Furthermore, a STA can play multiple roles in a single sensing procedure~\cite{case} as shown in Fig~\ref{case}.
\begin{itemize}
\item In the first case, the sensing initiator is the sensing
receiver, which directly obtains measurements by itself
using PPDUs transmitted by the sensing responder.
\item The sensing initiator in the second case is the sensing transmitter, which transmits PPDUs and performs the sensing function using feedback from the sensing responder.
\item In the third case, the sensing initiator functions as both a sensing transmitter and a sensing receiver, allowing it to obtain uplink measurements by receiving PPDUs and downlink measurements through feedback.
\item  In the last case the sensing initiator is neither a sensing transmitter nor a sensing receiver,the sensing responder feeds back the measurements obtained by other means to the sensing initiator, thus allowing the sensing initiator to obtain the measurements without sensing PPDUs.
\end{itemize}
The main contribution of the IEEE 802.11bf amendment will be the specification of procedures that allow WLAN sensing applications to obtain measurement results in a reliable and efficient manner. Specifically, IEEE 802.11bf can provide a service that enables a STA to obtain sensing measurements of the channel between two or more STAs and/or the channel between a receive antenna and a transmit antenna of a STA through the sensing procedure.
\\a WLAN sensing procedure typically contains five phases, namely the sensing session setup, sensing measurement setup, sensing
measurement instance, sensing measurement termination, and
sensing session termination. It should be noted that IEEE 802.11bf standarization is still ongoing. 
In the following the IEEE WLAN sensing procedures that has been developed by TGbf.
\begin{enumerate}
\item \textbf{Sensing Session Setup:} During this stage, the sensing initiator establishes a sensing session with the sensing responder, and sensing-related capabilities are exchanged between them. Multiple sensing sessions can coexist, with each sensing session uniquely identified by the MAC address of the STA establishing the sensing session. Furthermore, multiple sensing sessions can be maintained by the same sensing initiator to meet the WLAN sensing procedure requirements.
\item \textbf{Sensing Measurement Setup:} Sensing Measurement Setup enables a sensing initiator and a sensing responder to exchange and agree on operational attributes associated with a sensing measurement instance, such as the role of the STA, the type of measurement report, and other operational parameters. Measurement setups with different sets of operational attributes are assigned different Measurement Setup IDs to identify a specific set of operational attributes.
\item \textbf{Sensing Measurement Instance:} Sensing measurements are performed in the sensing measurement instance. The Measurement Instance IDs can be used to identify different sensing measurement instances.
\item \textbf{Sensing Measurement Termination:} The corresponding sensing measurement setups are terminated during sensing measurement termination. The sensing initiator and sensing responder release the resources allocated to store the sensing measurement set.
\item \textbf{Sensing Session Termination:} In this stage the STAs stop performing measurements and terminate the sensing session.
\end{enumerate}
\section{Channel and Target Models for Sensing}
In contrast to previous channel models for standards operating at sub-7 GHz and 60 GHz, such as IEEE 802.11ax and IEEE 802.11ay, respectively, which entirely focused on communication characteristics in specific frequency bands, the new channel model for WLAN sensing should primarily focus on providing accurate space-time characteristics of the propagation channel arising from the devices and free moving targets in both sub-7 GHz and 60 GHz bands. For this reason, the 802.11bf channel model document~\cite{channelmodel} presented two preliminary channel models focused on WLAN sensing, namely the ray tracing and data-driven hybrid channel models.
\\The 802.11bf channel model document describes channel models for Wi-Fi sensing systems based on ray tracing results and other possible approaches, such as measurement campaigns. The channel modeling is intended to aid in the standardization of WLAN Sensing. The document provides channel models that account for the non-stationarity features of the propagation channel caused by device-free moving targets at sub-7GHz and 60GHz. The channel models enable the generation of a channel realization that comprises the spatial, temporal, and amplitude properties of all rays in the channel realization. For both a transmitter and a receiver, the spatial features of rays include azimuth and elevation angles. Reference antenna models that may be potentially applied to the generated space-time channel realizations in the channel model are described. Directional and omni-directional antenna models are proposed to be used together with the channel models.
\subsection{Ray Tracing-Based Channel Model}
A raytracing based channel model is a deterministic model that processes channel realization using a ray-tracing technique 
By considering the transmitted signal as a particle, the ray-tracing technique is a numerical computational electromagnetics technique that uses a computer program to generate estimates for multipath characteristics such as path loss, angle of arrival (AOA), angle of departure (AOD), and time delays. The ray-tracing-based channel model includes as input scenarios, targets, and antenna configurations to generate all possible time-variant rays for each TX-RX pair in various simulation frames. Following this stage, the channel impulse response is created and ready for simulation.
\\the ray characteristics can be derived from the geometrical properties of a ray, The delay is defined as d/c, i.e. the ratio of the path length d(m) and the speed of light c (m/s). The free space path loss scales with the path length, The path loss in dB can be computed using the Friis transmission loss equation:
\begin{equation}
 path loss=20 log(4\pi d/\lambda)
  \label{(2)}
 \end{equation}
\\The phase of a ray is assumed to be rotated by 180 degrees for every reflection.

\subsection{Data-Driven Hybrid Channel Model}
The rays between TX and RX in the data driven hybrid channel model (DDHC) are divided into target-related rays and target-unrelated rays, with the target-related rays generated using the previously stated ray-tracing method and the target-unrelated rays generated by using an auto regressive model of existing standardized channel models (IEEE 802.11ax for sub-7 GHz and IEEE 802.11ay for 60 GHz) The final DDHC Model can be generated by using the real dataset acquired from experiments to better resemble the real channel model in a specific scenario. The DDHC Model can be utilized to help in sensing uncertainty modeling. Three sorts of causes contribute to sensing uncertainty: unpredictable reflections from the wall and random scatters, irregular target movement, and receiver noise.
\\The target-unrelated rays are generated by combining the autoregressive model and the channel model in 11ax (for sub 7GHz) or 11ay (for 60GHz). The auto-regressive model is given in equation:

\begin{equation}
\sigma=
\begin{cases}
\gamma (t) &  0\<t\<T_0 \\
\rho \sigma_i (t-T_0 )+(1-\rho)\gamma(t) & t>T_0
\
\end{cases}
  \label{(1)}
 \end{equation}
\\where $\sigma$i is the target-unrelated rays, $\gamma$ is the 11ax/ay channel model, and T0 and $\rho$  are the coherent time hyper-parameters, respectively.
\\for 11ay:
$$\sum_{i=0}^{N_rays-1} (A^i U_i^c*h (V_i^c*h )^H \delta(t-t_i ))$$
\begin{equation}
 \sum_{i=0}^{N_rays-1} (A^i U_i^c*h (V_i^c*h )^H \delta(t-t_i ))
  \label{(3)}
 \end{equation}
\\where Nrays is the total number of rays, $A^i$ is the amplitude of the ith ray, $U_i^c*h$ and $V_i^c*h$ are the channel phasor vectors of the ith ray, ti is the time delay of the ith ray.
\subsection{Global Human Walking Model}
Since human sensing is considered in the most of IEEE 802.11bf use cases, precise modeling of these applications is required. Boulic and Thalmann created a global human walk model based on empirical mathematical parameters derived from biomechanical experimental data~\cite{boulic1990global}, which has been a subject of study in a range of domains, including biomedical engineering. This model is an averaging human walking model without information on personalized motion characteristics since it is built on averaging parameters from experimental measurements. A body segment's 3-D orientation is established by identifying the 17 joint points and 16 segments using body reference coordinates centered at the origin of the spine.
\\The angle rotation matrix is used to determine the locations of the 17 joint points at each frame time based on the flexing angle functions and translations of the joint points specified by the model of biomechanical experimental data. The 3-D trajectories of these joint locations are generated by carefully adjusting the flexing and translation. Animating a model using data from a series of reference points throughout time has proved the model's validity and it demonstrates that it is capable of producing a realistic walking human model~\cite{chen2019micro}.
\begin{figure}[ht]
    \centering
    \includegraphics[scale =1]{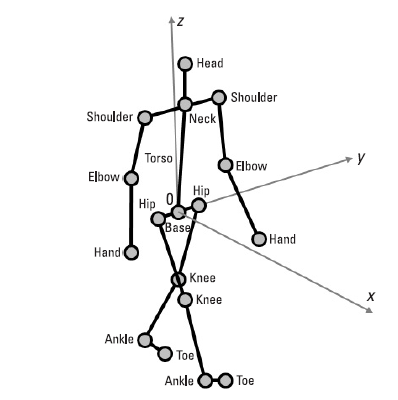}
    \label{}
    \caption{Reference points of the human model~\cite{chen2019micro}.}
\end{figure}
\\The global walking model has been successfully used in the design of radar systems~\cite{chen2019micro}, since The trajectories of various walking human body parts and the related radial velocities reveal that the model's micro-Doppler signature pattern is similar to the micro-Doppler signature pattern of radar backscattering from a walking person.
\section{Threshold Sensing}
One of the greatest considerations in WLAN sensing is sensing measurement and reporting, especially when the device that takes the measurements is not a sensing processor. The threshold sensing approach has been offered as a way to reduce the amount of feedback for reporting the data.
\\As discussed previously, a sensing initiator can act as a transmitter, receiver, both, or neither, but a sensing responder can act as a transmitter, receiver, or both. It should be noted that if a sensing initiator is also a sensing receiver, no feedback is required since the initiator may collect the measurements directly and use them to determine the sensing result. On the other hand, feedback is necessary to get CSI measurement from the sensing receiver if the sensing initiator is a sensing transmitter. The measurement and reporting technique for threshold-based sensing has been detailed in~\cite{threshold} as an optional procedure.
\\The frequency of feedback is determined by the use case. Some use cases, such as intruder detection, need frequent CSI feedback, yet, the majority of the feedback may be strongly linked over time. As a result, the sensing receiver is not required to constantly feedback the CSI. The receiver can feed back only when the CSI variation becomes large. In the reporting step of the method, a threshold is employed for this purpose. the transmitter reconstruct the missing CSI measurement using interpolation techniques.
\\The selection process for the CSI variation value depends on the implementation, however it must respect the following guidelines: The CSI variation value must be within the closed interval [0, 1]. When the CSI variation result is equal to 0, it means that the CSI variations are less than a particular limit and remain constant throughout all future measurement instances using the same measurement setup. The CSI variation value of 1 denotes situations where the CSI variations are greater than a fixed value and remain constant throughout all future measurement instances using the same measurement setup.
\\The CSI variation might be measured by the time-reversal resonating strength (TRRS)~\cite{wu2015time}, which is the maximum amplitude of the entries of the cross-correlation between two complex CIRs. Since it takes the greatest value of the correlation coefficients, this approach is more resilient than traditional correlation coefficients.

\chapter{Mm-Wave WiFi Physical Layer Models}
This chapter provides the 60 GHz Wi-Fi physical layer models which will be used for sensing evaluation.
\section{802.11ay CSI Based Sensing}
To enable sensing applications on a communication system,the system model under consideration is an IEEE 802.11ay single carrier (SC) system. the sensing system is using the CSI for sensing and a correlation receiver. The CSI, which is already present in a standard IEEE 802.11ay system, is used to sense the environment by tracking channel variations over time. Assuming a static transmitter and receiver, a change in the environment, such as a moving target, causes a CSI variation.
\\In this model the receiver is assumed to be the sensing session's initiator, i.e., the node requesting sensing information and The transmitter continuously sends IEEE 802.11ay packets. before we get into the sensing operations. reviewing the EDMG packet format is necessary.
\section{EDMG Packet Format}
For single-user transmission, the packet formats defined in IEEE 802.11ad and IEEE 802.11ay are referred to as directional multi-gigabit (DMG) PPDU and EDMG PPDU, respectively. 
in a DMG packet, The short training field (STF) enables packet detection, gain control, and carrier frequency and timing acquisition. The channel estimation field (CEF) is used for channel estimation as the name implies. The header contains information needed to demodulate the packet, such as the modulation coding scheme (MCS) used, while the data field contains the payload data and possible padding, and the automatic gain control (AGC) enables the receiver to re-adjust its AGC setting before processing the training (TRN) field. 
To ensure coexistence with DMG stations, DMG stations can detect the first portion of an EDMG packet.
L-STF, L-CEF, and L-Header have the same definitions and applications as the STF, CEF, and header fields of DMG packets, respectively. Only EDMG stations recognize the second portion of an EDMG PPDU. The EDMG Header- A field that contains information needed to interpret the packet, such as bandwidth and the number of spatial streams.
\begin{figure}[ht]
    \centering
     \begin{subfigure}[b]{0.9\textwidth}
         \includegraphics[width=\textwidth]{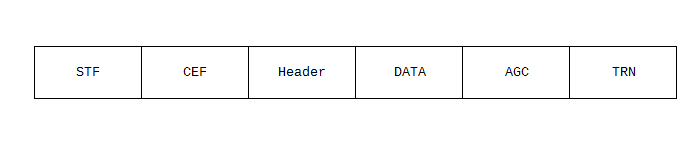}
         \caption{}
         \label{}
     \end{subfigure}
     \hfill \begin{subfigure}[b]{0.9\textwidth}
         \includegraphics[width=\textwidth]{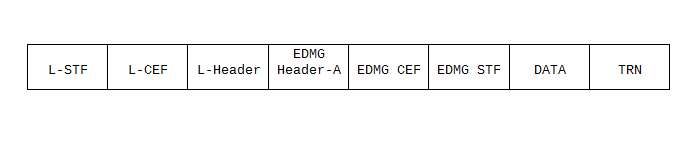}
         \caption{}
         \label{}
     \end{subfigure}
    \label{}
    \caption{(a)DMG packet format (b)EDMG packet format.}
\end{figure}
The EDMG-STF and EDMG-CEF fields enables EDMG stations to estimate various signal parameters and the channel when channel bonding, channel aggregation, and/or MIMO are used.The EDMG-Header-B is only
included in MU-MIMO packets.
\\A single packet format is defined for the three EDMG PHY modes: SC, OFDM, and control. Not all fields are transmitted in an EDMG packet. fields are included depending on whether the packet is used for single channel or channel bonding operation, for SISO or MIMO transmission, and if it is used for beamforming training or tracking procedure~\cite{da2018analysis}
\\since we reviewed the EDMG packet format the sensing operations are carried out in these steps. First, the frame is detected and synchronized by locating the peak of the correlation between the received signal and the known STF pilots, then the coarse frequency offset is then determined by comparing the phase changes to the STF correlation peaks. joint frequency offset and channel estimation are performed by correlating the received CEF with the known CEF. The channel estimated using the preamble can be seen as echoes from the targets and the environment from a remote sensing perspective. The echoes from the target multi-path components (MPCs) have a delay proportional to the bi-static distance.
\\as IEEE 802.11ay packets are continuously sent over time, the communication receiver reports the variation of the CSI to the sensing processor within the receiver,the sensing processor can construct the radar data matrix by collecting the estimated channel impulse response (CIR) at each packet reception in a 2-dimensional matrix referred to as the radar data matrix~\cite{blandino2022tools}, this procedure will be demonstrated in the next chapter.
\section{802.11bf Multi-Static PPDU}
IEEE 802.11bf enables the possibility to re-use IEEE 802.11ay beam refinement protocol (BRP) frames for sensing purposes as well introducing a dedicated multi-static PPDU.
\\A multi-static EDMG sensing PPDU is an EDMG BRP PPDU with some modifications, such as the removal of the data and EDMG-CEF fields from the PPDU structure, and the addition of a Sync field after the EMDG-A Header. An EDMG Multi-Static Sensing PPDU enables sensing by multiple STA using the same PPDU, These are dedicated sensing packets, which are only used for sensing tasks, also known as active sensing. A STA that is participating in an EDMG Multi-static Sensing Instance as a receiver may ignore the L-STF, L-CEF, L-Header and EDMG-Header and use its intended Sync Subfield for synchronization. a sequence of Golay sequences can be sent to train the receiver and transmitter antenna in the TRN fields~\cite{multistatic} which we discuss in details below. 
\\\begin{figure}[h]
    \centering
    \begin{subfigure}[b]{1\textwidth}
         \includegraphics[width=\textwidth]{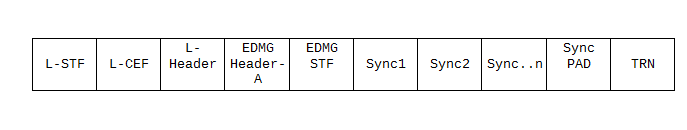}
     \end{subfigure}
    \label{}
    \caption{EDMG multi-static PPDU.}
\end{figure}
\subsection{Training Field (TRN) Format}
The TRN field, which is appended to packets used in a beam refinement protocol, enables transmit and receive beamforming training. BRP is a process by which a station can improve its antenna configuration for transmission and/or reception. The TRN field in IEEE 802.11ay was redesigned to increase efficiency and make it configurable based on the characteristics of the specific beamforming training procedure being executed.
\\The TRN subfield, which is composed of Golay complementary sequences, is the TRN field's basic unit. A TRN-Unit is formed by concatenating a variable number of TRN subfields. The TRN field is composed of a variable number of TRN-Units, which is defined by the EDMG TRN Length parameter.
\\All TRN subfields are transmitted with the same AWV as the data field in a BRP procedure used for receiver training. such packets are referred to as EDMG BRP-RX packets,AWV is a vector of weights that describes the excitation (amplitude and phase) for each element of an antenna array. When receiving different TRN subfields, the receiver can switch AWVs and determine an improved antenna configuration setting.
\begin{figure}[h]
    \centering
    \includegraphics[scale=0.4]{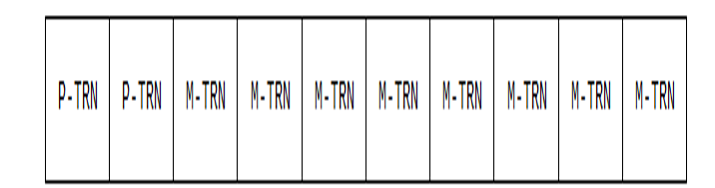}
    \label{}
    \caption{EDMG TRN field structure for unit p=2, unit M=8, unit N=1.}
\end{figure}
\\In a BRP procedure used for transmitter training, the transmitter transmits the TRN field with different AWVs, The format and length of a TRN-Unit used for transmitting beamforming training also referred to as EDMG BRP-TX packets, are defined by three parameters: EDMG TRN-Unit P, EDMG TRN-Unit M, and EDMG TRN-Unit N, which are referred to as P, M, and N TRN. The first P TRN subfields in a TRN-Unit are transmitted with the same AWV as the data field. Therefore, such TRN subfields may be used by the receiver to maintain synchronization and estimate the channel. in the transmission of the remaining M TRN subfields of a TRN-Unit The transmitter may change the AWV at the beginning of each TRN subfield To improve the beamforming training process, the last M TRN subfields of a TRN-Unit may be transmitted with more than one consecutive AWV. N is the number of consecutive TRN subfields transmitted with the same AWV~\cite{ghasempour2017ieee}.
\\As demonstrated above, when using IEEE 802.11ay PPDUs for sensing, the packets are sent continuously over time with omni antennas, and the sensing processor can build the radar data matrix, collecting the estimated CSI at each packet reception in a 2-dimensional matrix, while when using a multi-static PPDU and  Assuming directional communication, the system precodes the TRN fields and The estimated CSI from the TRN is collected in a 3-dimensional matrix including information about the angular domain.

\chapter{Sensing Data Acquisition and Signal Processing Framework}
This chapter introduces the used sensing signal processing method which mainly used in digital pulse Doppler radar.
\\The sensing system transmits a perfectly known signal into free space, with the expectation that a version of the transmitted signal will be reflected back from the target to the sensing receiver. The propagation effects change the frequency and phase of the sensing signal as it travels from the transmitter to the target and back (in the case of mono-static sensing). The total input signal at the receiver's input is composed of this modified version of the transmitted signal and an additive noise signal. The receiver's primary purpose is to process the noisy received signal using signal processing in order to extract target properties such as range, radial velocity, etc. These operations are carried out by the sensing receiver using signal processing techniques to enhance target detection. By using the differences between the transmitted and received signals, various signal processing techniques and/or algorithms may be used to extract target properties. One of the most common of these signal processing techniques which can be easily used with CSI is correlation.
\section{Correlation Receiver}
Correlation in signal processing reveals how similar two signals are to one another. The correlation receiver subsystem estimates the unknown target properties by assessing how similar the returning target echoes are to the original transmitted signal, where cross correlation is done between the receiving signal which represents the target echo and the transmitting signal, the correlation peak appears at the time delay corresponding to the target range. However, if the target is moving a Doppler shift would be introduced to the echo signal, this Doppler shift will distort the output of the correlation where the correlation peak will no longer be optimal, which concludes that the Doppler information of the target cannot be detected with a single transmission using the correlation receiver since the correlation only provides information about the delay, also In actual use, the sensing receiver receives both the echo signal from the moving target and the echo signal from stationary objects. The echo signals due to stationary objects are called clutters. For this purpose the sensing receiver employs the Doppler Effect principle to discriminate non-stationary targets from stationary ones~\cite{mahafza2005radar}. 
\section{Doppler Processing}
The term Doppler processing refers to the filtering or spectral analysis of a signal received from a fixed range over a period of time corresponding to several pulses. The spectrum of a signal received from a single range consists of noise clutter and one or more target signals, and in many situations, the amplitude of the target signal is above the noise (SNR $>>$ 1) but below the clutter (signal to clutter ratio $<<$ 1), Since real target detection is not possible in this situation, Doppler processing is utilized to separate the target and clutter signals in the frequency domain by filtering the clutter while retaining the target returns.
\\doppler effect of the target means that If the target moves in the direction of the radar the frequency of the received signal will increase. Similarly, when the target moves away from the Radar, the frequency of the received signal decreases.
\\a widely used technique in sensing processing to estimate the range and velocity of moving targets is range Doppler map, a range Doppler map can be generated from a radar data matrix using signal processing techniques such as discrete Fourier transforms (DFT) or high resolution algorithms such as space-alternating generalized expectation maximization(SAGE) or Multiple signal Classification (MUSIC).
\subsection{Radar Data Matrix}
A radar data matrix has two time dimensions: a fast time and a slow time. Each slow time instance in a radar data matrix has one received sample that was gathered during a single pulse. The identical data gathering procedure is repeated over a number of pulses with pulse interval on the slow time axis. The target's delay can be determined from the transmitter and the target and receiver using correlations along the fast time axis.
\\The fast time signal bandwidth, which is typically equal to the transmitted pulse bandwidth, is at least equal to the sampling rate in the fast time or range dimension. For a sensing system that use a pulse of length t, The slow-time or pulse number dimension is sampled at the pulse repetition interval (PRI) T of the radar. Therefore, the sampling rate in this dimension is the pulse repetition frequency (PRF).
\begin{figure}[ht]
    \centering
    \includegraphics[scale=0.3]{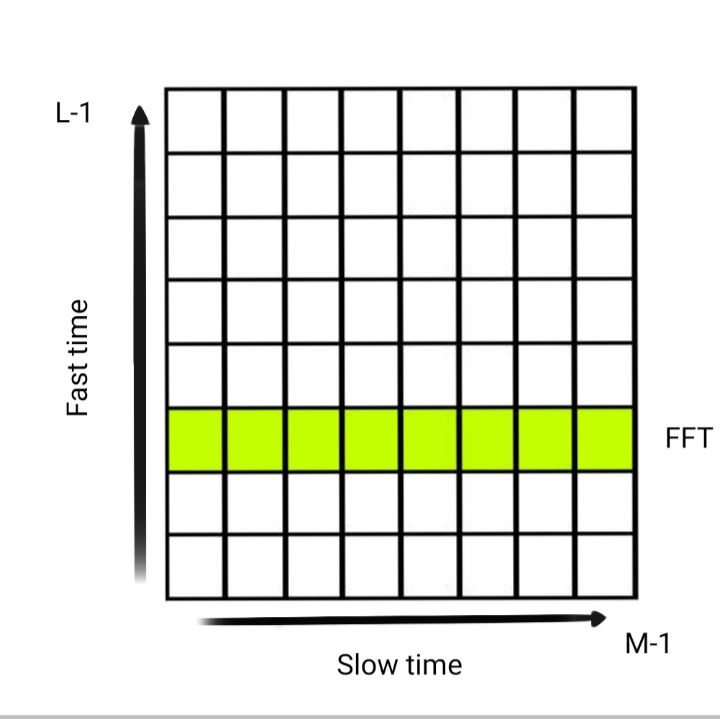}
    \label{}
    \caption{Representation of the radar data matrix.}
\end{figure}
\\The matrix's rows each reflect a series of measurement taken over M successive pulses from the same range bin. The coherent processing interval (CPI) is the total amount of time MT that the data matrix represents. a CPI of data is typically gathered utilizing a constant PRI and constant frequency. The  CPI is used to refer to both the matrix of data and the time necessary to acquire it~\cite{richards2014fundamentals}.
\\The radar can recover the spatial features of the target by utilizing an antenna array with beamforming applied to the received channels. In this case, the data metrics from each channel are stacked to form a data cube, to provide information about angular properties of the target beyond the delay and Doppler shift.
\begin{figure}[ht]
    \centering
    \includegraphics[scale=1]{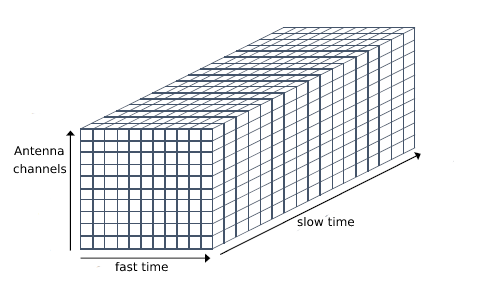}
    \label{}
    \caption{Radar data cube.}
\end{figure}
\subsection{Short Time Fourier Transform}
Short-time Fourier transform (STFT) is an improved version of Fourier transform that is used for transforming non-stationary time-series signals into frequency domain (signals whose frequency components change over time). STFT extracts time-series signal segments using a fixed-length window and performs the Fourier transform to each extracted segment of the signal, resulting in time-localized frequency information for the signal. The typical Fourier transform, on the other hand, considers the entire signal and returns frequency information that is averaged over the entire time domain, losing information about the time when these frequencies occurred in the time-series signal~\cite{kehtarnavaz2011digital}.

\begin{equation}
  X_{STFT}=\sum_{k=1}^{L-1} X[k]*g[k-m]*e^{-j2\pi*nk/L}
  \label{(1)}
 \end{equation}
\\where x[k] denotes a signal and g[k] denotes an L-point window function. The STFT of x[k] can be interpreted as the Fourier transform of the product x[k]*g[k-m].
\\STFT uses windowing technique which is discussed further below.

\subsubsection{Window Type}
Depending on the signal, you can use several different sorts of window functions. To understand how a certain window impacts the frequency spectrum, A real-world plot of a window reveals that its frequency characteristic is a continuous spectrum with a main lobe and multiple side lobes. The primary lobe of the time-domain signal is centered at each frequency component, while the side lobes approach zero. The height of the side lobes reveals how the windowing function affects frequencies around the main lobes.
\\The sinusoidal form is shared by the Hamming and Hann window functions. Both windows provide a wide peak with narrow side lobes. However, the Hann window touches zero at both ends eliminating all discontinuity. Since the Hamming window does not quite approach zero, there is still a little discontinuity in the signal. Due to this difference, the Hamming window cancels the nearest side lobe better but fails to cancel any others. These window functions are useful for noise measurements where more frequency resolution than other windows is desired but moderate side lobes are not an issue.
\begin{figure}[ht]
     \centering
     \begin{subfigure}[b]{0.32\textwidth}
         \includegraphics[width=\textwidth]{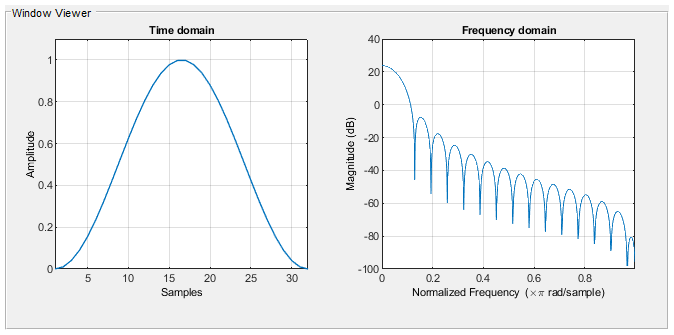}
         \caption{}
         \label{}
     \end{subfigure}
     \hfill
     \begin{subfigure}[b]{0.32\textwidth}
         \includegraphics[width=\textwidth]{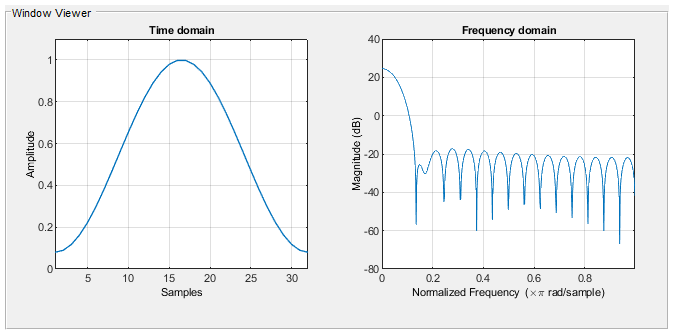}
         \caption{}
         \label{}
     \end{subfigure}
     \hfill
\begin{subfigure}[b]{0.32\textwidth}
         \includegraphics[width=\textwidth]{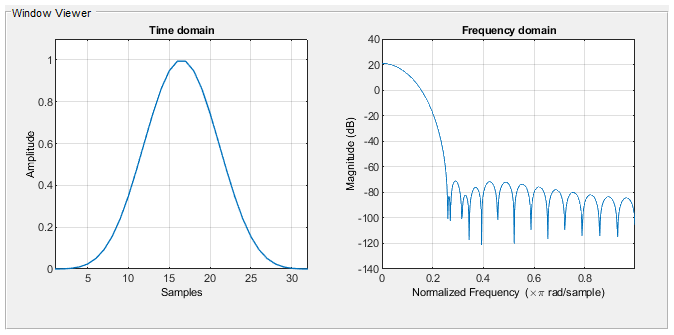}
         \caption{}
         \label{}
     \end{subfigure}
        \caption{FFT window types (a)hann window (b)hamming window  (c)blackman-harris window. }
        \label{}   
\end{figure}
\\The Blackman-Harris window is similar to the Hamming and Hann windows. The resulting spectrum features a wide peak but good side lobe compression. The Blackman-Harris is an excellent general-purpose window with a high side lobe rejection and a moderately wide main lobe.
\subsubsection{Window Length}
Even while STFT attempts to maintain the signal's time-localized frequency information, there is still a trade-off between time and frequency resolution. This trade-off emerges since STFT transforms the time-series signal into frequency domain using a fixed-length window. The longer the window employed, the better the frequency resolution and the lower the time resolution. On the contrary, the shorter the length of the window employed, the lower the frequency resolution and the greater the time resolution. Since STFT employs a fixed-length window, the frequency resolution of the STFT is the same across all positions in the spectrogram.
\subsubsection{Window Overlap}
When processing data to determine the frequency content of a signal, we must feed a block of time data through the Fast Fourier Transform. Overlap happens when two contiguous observation time blocks overlap and include the same time data, as the name implies. Overlap is expressed as a percentage and refers to the proportion of the observation time that overlaps the adjoining observation time block. Overlap indicates that the time data in the overlapped area corresponds to more than one FFT computation. Due to the overlapping observation times, less of the original time data is used for a certain number of observations.
\\When investigating the frequency content of a measured signal to do more than one FFT computation on our time data, the influence of the window functions used for each observation must be considered. If the time data is analyzed with the observation times one after the other with no overlap, there will be portions of the original measurement that are zeroed-out by the window and will not be utilized in any of the FFT computations. When utilized correctly, overlap can assist overcome these window effects, for example, The use of 90 percent Overlap shows more accurate amplitude levels across all frequencies and moments in time.
\subsection{Range Doppler Map}
After constructing the radar data matrix, sensing signal processing is performed on the slow time dimension of the matrix. The range and radial velocity can then be calculated from the delay and the Doppler shift respectively. this results in a range Doppler map.
\\The aim of sensing applications is to detect targets by analyzing the impact of the target on wireless signal propagation over time. However, in a rich scattering environment, the received radio signal may contain unwanted clutters originating from static objects in the environment, in addition to the signal scattered from the target. For various sensing scenarios and deployments, several clutter reduction methods have been developed To reduce the clutter, for example,Huang et al~\cite{huang2021indoor} designed the scheme to simply remove the direct current (DC) component in the range-Doppler map from each range bin in order to suppress the clutter.
\subsubsection{Micro-Doppler}
Typical radar data is displayed in a number of picture types, starting with the range-Doppler map, to images of radar tracks, which are typically a series of dots, and sometimes presenting the spectrogram of the tracked target. This procedure enables the operator to examine the specifics of each track in the spectrogram while maintaining a very rapid update rate for target positions in the range-Doppler map. A spectrogram is rarely required for vehicular targets, but for human or animal targets, spectral information can assist in the classification and analysis of detection and tracks~\cite{tahmoush2011time}.
\\A micro doppler plot is a time-integrated range-Doppler map that displays the  properties of targets in radar images, allowing an operator to distinguish different target types and actions performed by the targets. A combination of range-Doppler maps over time result in a spectrogram like characterization of Doppler while retaining range information. These are compiled from the range-Doppler maps by taking the maximum value for each pixel over a time range.

\chapter{Bi-Static Sensing Performance Evaluation}
 In this chapter the performance of the physical layer models discussed in chapter 3 are investigated as a sensing technology using the software provided by~\cite{blandino2022tools}. 
\section{Simulation Setup}
\begin{figure}[h]
    \centering
    \includegraphics[scale = 0.8]{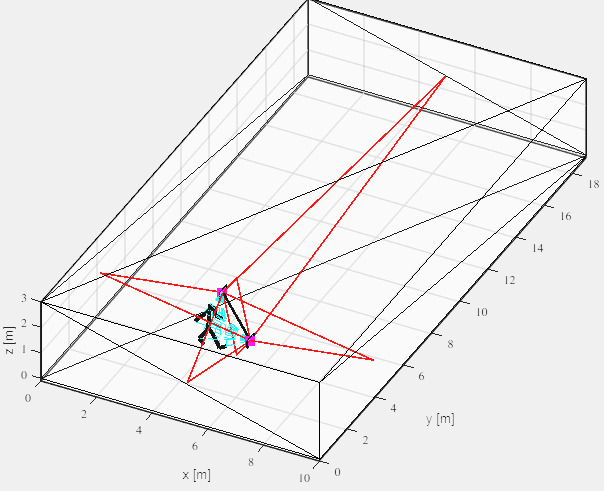}
    \caption{Empty room environment with a SISO link and a human target.}
    \label{ray}
\end{figure}

creating an environment model for the scenario is the first step for WLAN sensing evaluation. In this scenario, an empty rectangle room with the dimensions 19x10x3m3 includes two WLAN nodes and a single human target moving along a predefined trajectory. The channel is a deterministic raytracing channel model adapted to the locations of the transmitter and receiver, the environment, and the mobility of the target. the model is shown in Fig. ~\ref{ray}
\\The transmitter is located at (4,5,1.5) while the receiver is at (6,3,1.5) In the given empty room map, the Q-D software is set to create 768 channel realizations. The whole duration of motion is considered to be 1.3 seconds. the carrier frequency of the simulation is assumed to be 60 GHz, and the time sampling of the channel is 1.69ms.
\\the human target model is generated using the software provided by~\cite{chen2019micro}, The simulation is configured to build an array of the coordinates for various joints in a walking human model, with the array size set to match the sampling of the deterministic channel. The walking human model's height is set to be 1.8 meters. and its trajectory is from (4,4) to (5,4).the software simulates the Doppler frequency shift to determine whether it is similar to a real human target micro-Doppler.
\section{System Evaluation Settings}
The system under consideration is a SISO EDMG SC transmission. The modulation coding scheme index is 12, which means that the SC blocks contain $pi$/2 16-QAM symbols with a coding rate of 1/2. The packet repetition frequency is 590 Hz, with a system bandwidth of 1.76 GHz and an SNR of 20 dB. A Blackman-Harris window is used to pre-filter the data before performing the Doppler FFT.
\\When modeling directional propagation, the antenna pattern of the antenna steering direction defined by the AWV should be considered. To study the effect of the AWV on performance, a collection of angular codebooks that provide the array steering vectors and antenna wave vectors will be tested. The antenna pattern features a single main beam that is directed in a certain direction. The entries in the angle-based codebook specify the phased antenna array steering directions. The codebooks offered are for various sizes:
\begin{itemize}
    \item 2X2: 2 vertical antennas and 2 horizontal antennas. This codebook includes 25 directions; a combination of 5 azimuth and 5 elevation angles.
    \\Azimuth:[0 45 90 270 315]
    \\Elevation:[0 45 90 135 180]
    \item 2X8: 2 vertical antennas and 8 horizontal antennas. This codebook includes 85 directions, combination of 17 azimuth and 5 elevation angles.
    \\Azimuth:[0 11 23 34 45 56 68 79 90 270 281 292 304 315 326 337 349]
    \\Elevation:[0 45 90 135 180]
    \item 8X8: 8 vertical antennas and 8 horizontal antennas, This codebook includes 289 directions, combination of 17 azimuth and 17 elevation angles.
    \\Azimuth:[0 11 23 34 45 56 68 79 90 270 281 292 304 315 326 337 349]
    \\Elevation:[0 11 22 33 45 56 67 78 90 101 112 123 135 146 157 168 180]
\end{itemize}
\subsection{EDMG Sensing}
For this case, The micro-Doppler is created for several window settings, such as window size and window overlap, to investigate the influence of signal processing on the performance of sensing. the sensing processor is configured to do a 128-length Doppler FFT.
\begin{figure}[ht]
     \centering
     \begin{subfigure}[b]{0.495\textwidth}
         \includegraphics[width=\textwidth]{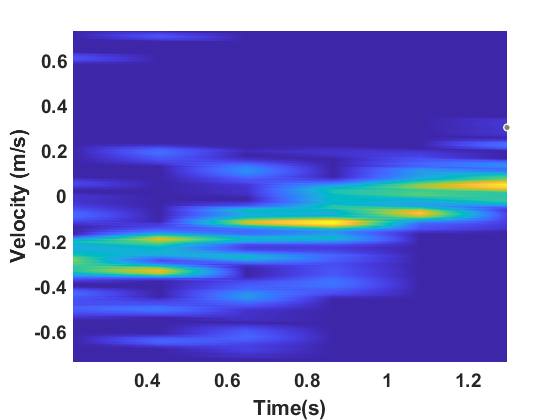}
         \caption{}
         \label{128}
     \end{subfigure}
     \hfill
     \begin{subfigure}[b]{0.495\textwidth}
         
         \includegraphics[width=\textwidth]{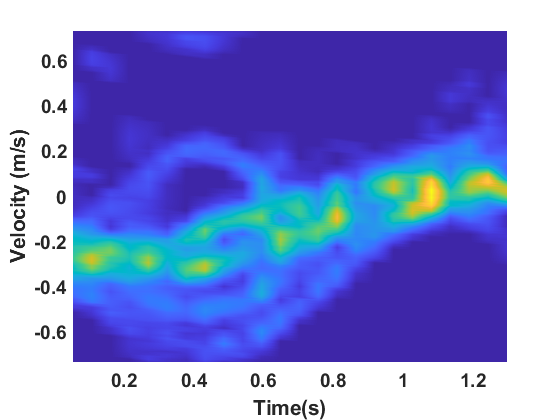}
         \caption{}
         \label{32}
     \end{subfigure}
     \hfill
     \begin{subfigure}[b]{0.495\textwidth}
         
         \includegraphics[width=\textwidth]{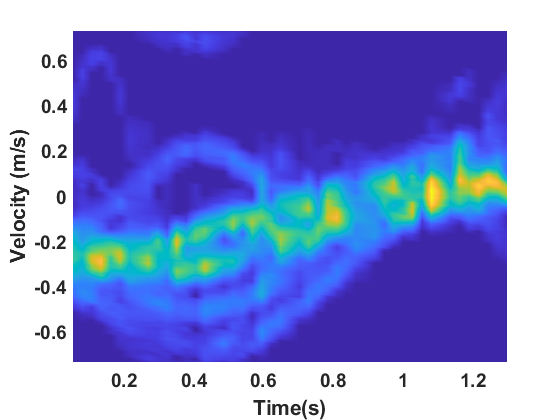}
         \caption{}
         \label{50}
     \end{subfigure}
      \hfill
\begin{subfigure}[b]{0.495\textwidth}
         \includegraphics[width=\textwidth]{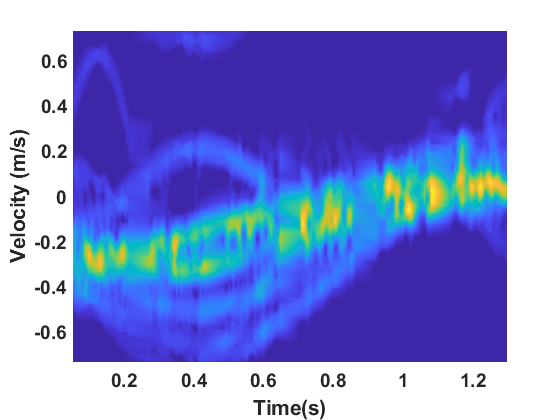}
         \caption{}
         \label{90}
     \end{subfigure}
        \caption{Micro-Doppler  with different window settings(a) FFT (b) STFT with 32 window length (c) STFT with 50\% overlap (d) STFT with 90\% overlap.}
        \label{w}   
\end{figure}
\\The micro-Doppler plots for different cases are shown in Fig.~\ref{w}. Fig.~\ref{128} shows the micro-Doppler plot for using FFT to process the data, this figure seems to not have the shape of a human walking micro-Doppler since it has very low time resolution due to the lack of windowing. Fig.~\ref{32} shows the micro-Doppler when using a 32 window size and it shows better time resolution as well as the oscillations of the radial velocity of different human joints over time.
\\Fig.~\ref{50} displays the micro-Doppler plot using 32 window size with a 50 percent window overlap, this plot differs from Fig.~\ref{32} in that it displays more information about the radial velocity and its oscillation for various joints, as well as more dots tracking the target. Fig.~\ref{90} displays a better tracking result with a 90 percent overlap by including more data to connect the dots at the target base and it presents a smoothing effect making the radial velocity tracking more apparent.
\subsection{Threshold Sensing}
\begin{figure}[h]
    \centering
     \begin{subfigure}[b]{0.495\textwidth}
    \includegraphics[width=\textwidth]{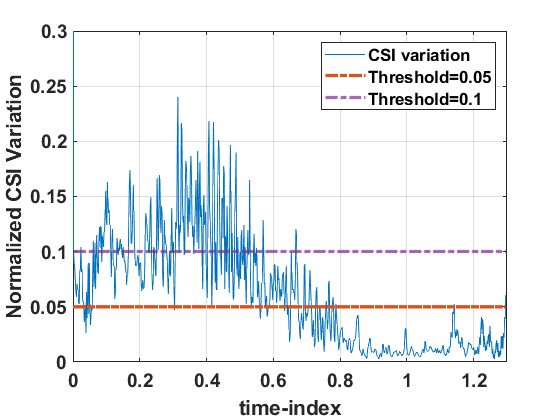}
    \caption{}
    \label{csi}
\end{subfigure}
\hfill
 \begin{subfigure}[b]{0.495\textwidth}
    \centering
    \includegraphics[width=\textwidth]{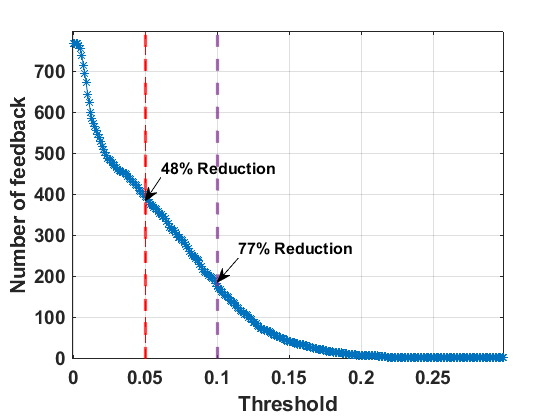}
    \caption{}
    \label{reduction}
    \end{subfigure}
    \caption{(a) Normalized CSI variation over time (b) Number of measurements need to be fed back versus threshold level}
\end{figure}

In this case, the TRRS metric is used to measure the CSI variation. Since the AP gets feedback data with irregular sampling, linear interpolation is used to reconstruct measurements with regular sampling. For signal processing, we use the same 32 STFT window size as in the previous EDMG sensing. The Normalized CSI variation over time is displayed in Fig.~\ref{csi} The figure includes the threshold levels used, Fig.~\ref{reduction} shows the number of measurements reported varying the threshold, from the figure it can be shown that using threshold levels of 0.05 and 0.1 results in a reduction of reporting CSI measurement by 48\% and 77\%, respectively.
\begin{figure}
     \centering
     \begin{subfigure}[b]{0.495\textwidth}
         \includegraphics[width=\textwidth]{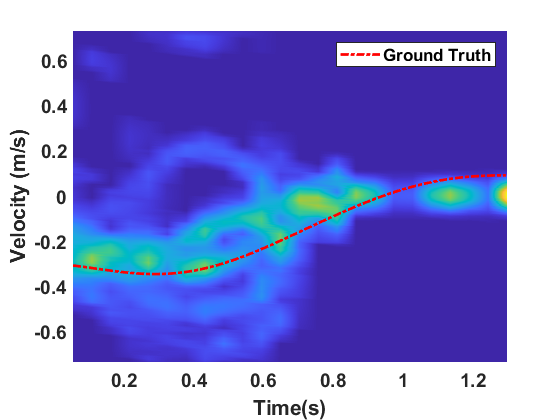}
         \caption{}
         \label{0.05}
     \end{subfigure}
     \hfill
\begin{subfigure}[b]{0.495\textwidth}
         \includegraphics[width=\textwidth]{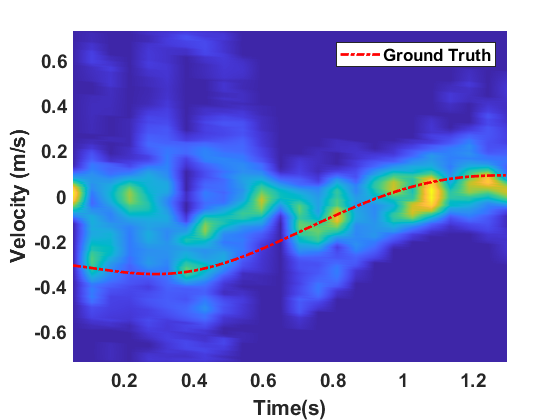}
         \caption{}
         \label{0.1}
     \end{subfigure}
        \caption{micro-Doppler plots with different  CSI threshold level (a) threshold=0.05 (b) threshold=0.1. }
        \label{threshold}   
\end{figure}
\\Fig.~\ref{threshold} displays the micro-Doppler plots for various threshold settings.comparing with the previous result without threshold, Fig.~\ref{0.05} shows that using a threshold of 0.05 the micro-Doppler is lacking a few radial velocity oscillations and provides a slight lower estimation accuracy. The result shown in Fig~\ref{0.1} with a threshold value of 0.1 appears to lose more data since it omits the oscillations of the human target joints and appears to provide lesser accurate results.
\\Overall, the utilization of threshold sensing is dependant on the use case. For our human walking scenario, the micro-Doppler results do not appear to be much worse when the number of CSI measurements is reduced, and it can still provide decent tracking results.
\subsection{Directional Sensing}
the channel considered in this case has the transmitter at (3,5,1.5) and the receiver at (4,7,1.5), with 600 channel realizations. This scenario considers a modulation coding scheme index of 2 with 40dB SNR, and a golay sequence length of 128. This scenario's sensing processing parameters are set to use a coherent process interval of 32 packets, FFT of length 64.
\subsubsection{Sensing Using TRN-R Packets}
The TRN architecture of a multi-static PPDU planned for IEEE 802.11bf is used to simulate the directional link. Using TRN-R packets to train a phased antenna array at the receiver while using an omni antenna at the transmitter. the training length for different antennas is configured considering that the TRN-R packets use 10 TRN subfields for beamforming training, so since the 2X2 antenna has 25 directions the training length is set to 3, the training length for the 2X8 antenna is set to 9, and the training length for the 8X8 antenna is set to 29.
\begin{figure}
     \centering
     \begin{subfigure}[b]{0.3\textwidth}
         \centering
         \includegraphics[width=\textwidth]{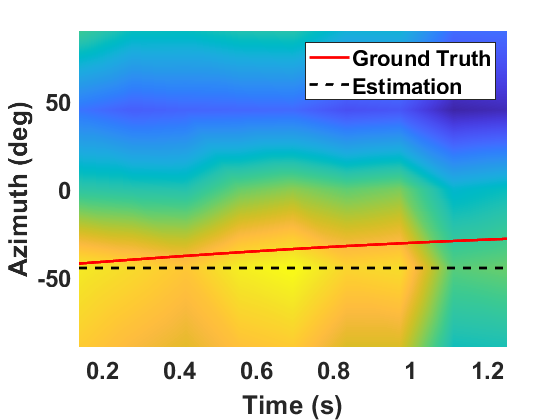}
         \caption{}
         \label{r2x2}
     \end{subfigure}
     \hfill
     \begin{subfigure}[b]{0.3\textwidth}
         \centering
         \includegraphics[width=\textwidth]{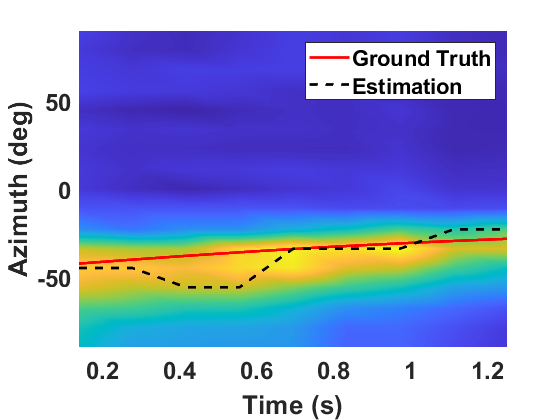}
         \caption{}
         \label{r2x8}
     \end{subfigure}
     \hfill
     \begin{subfigure}[b]{0.3\textwidth}
         \centering
         \includegraphics[width=\textwidth]{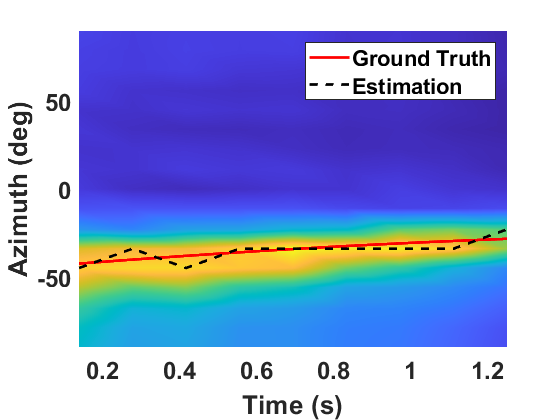}
         \caption{}
         \label{r8x8}
     \end{subfigure}
      \hfill
     \begin{subfigure}[b]{0.3\textwidth}
         \centering
         \includegraphics[width=\textwidth]{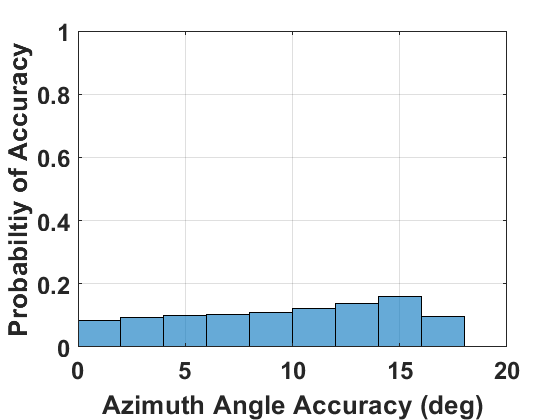}
         \caption{}
         \label{acr2x2}
     \end{subfigure}
     \hfill
     \begin{subfigure}[b]{0.3\textwidth}
         \centering
         \includegraphics[width=\textwidth]{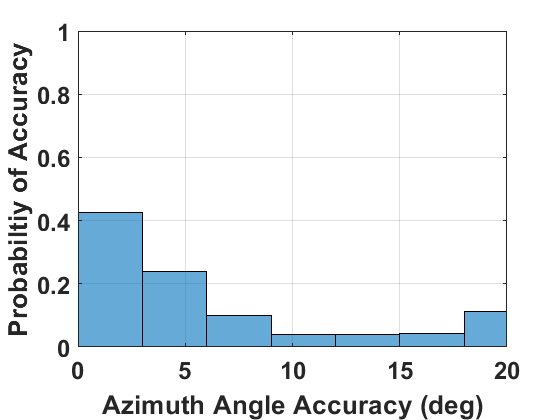}
         \caption{}
         \label{acr2x8}
     \end{subfigure}
     \hfill
      \begin{subfigure}[b]{0.3\textwidth}
         \centering
         \includegraphics[width=\textwidth]{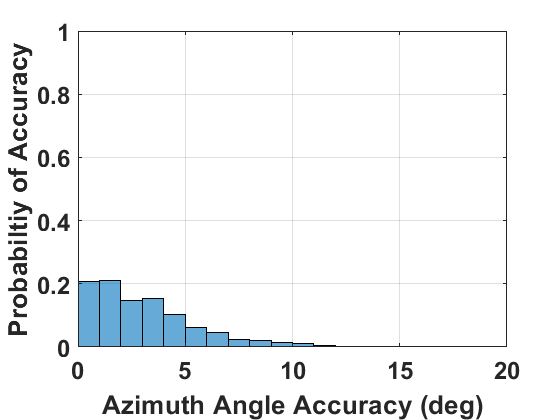}
         \caption{}
         \label{acr8x8}
     \end{subfigure}
        \caption{Azimuth angle estimation and accuracy for different antenna arrays using TRN-R packets (a) 2X2 estimation  (b) 2X8 estimation (c) 8X8 estimation (d) 2X2 accuracy histogram (e) 2X8 accuracy histogram  (f) 8X8 accuracy histogram.}
        \label{azr}   
\end{figure}
\\For various phased antenna arrays, the azimuth angle estimation and its associated accuracy are shown in Fig.~\ref{azr}. The figure shows that the estimation of the 2x2 antenna is far worst that the 2x8 and 8x8 antennas. This happens since the 2x2 antenna has a 5 azimuth steering angles while the 2x8 and 8x8 have 17 azimuth angles. the 2x8 and 8x8 antennas estimation performance are similar but it is clear that since the 8x8 antenna has more elevation Angles resulting in more directions, it provides more accurate results.
\subsubsection{Sensing Using TRN-T Packets}
In this scenario, the system is training a phased antenna array at the transmitter while using an omni antenna at the receiver using TRN-T packets. The TRN-T packet subfields are set as unit P=2, unit M=15, unit N=1.
\\it should be noted that the effect of changing TRN-T field parameters will not be considered since there effect is more relevant in real time scenario.
\\The training length for different antennas is configured considering the TRN-T packet field parameters, since the M field length is set to 15 we have 15+1 subfieds for beamforming training. Thus the 2X2 antenna training length is set to 2,and training length for the 2X8 antenna is set to 6, and the training length for the 8X8 antenna is set to 19.
\begin{figure}[ht]
     \centering
     \begin{subfigure}[b]{0.3\textwidth}
         \centering
         \includegraphics[width=\textwidth]{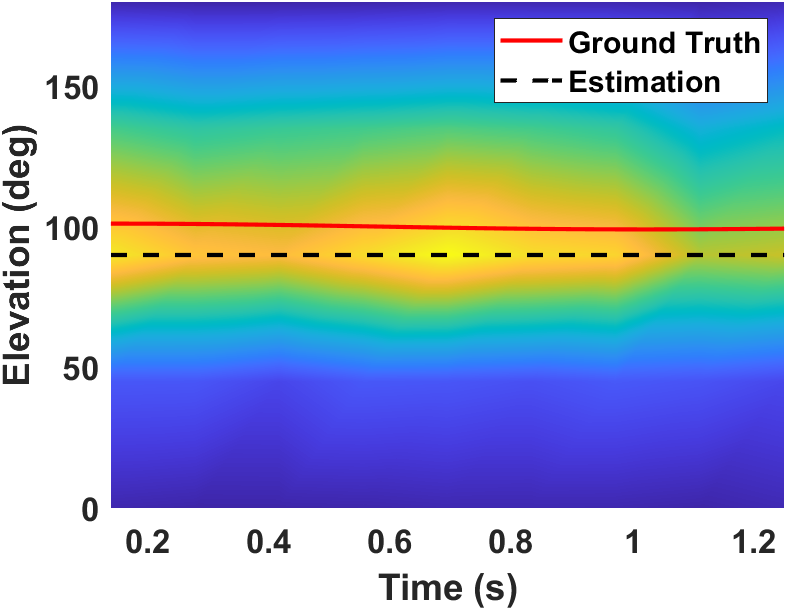}
         \caption{}
         \label{t2x2}
     \end{subfigure}
     \hfill
     \begin{subfigure}[b]{0.3\textwidth}
         \centering
         \includegraphics[width=\textwidth]{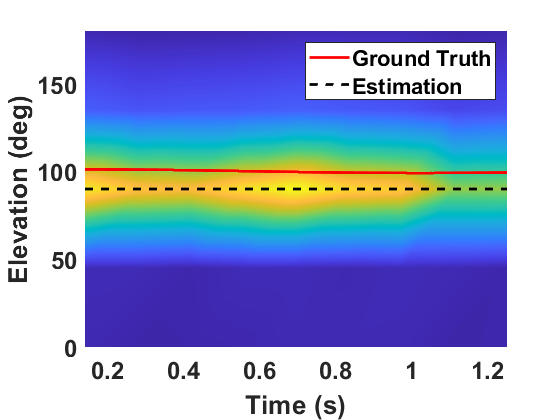}
         \caption{}
         \label{t2x8}
     \end{subfigure}
     \hfill
     \begin{subfigure}[b]{0.3\textwidth}
         \centering
         \includegraphics[width=\textwidth]{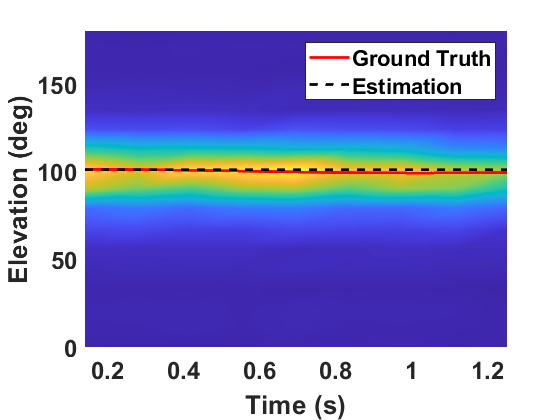}
         \caption{}
         \label{t8x8}
     \end{subfigure}
      \hfill
     \begin{subfigure}[b]{0.3\textwidth}
         \centering
         \includegraphics[width=\textwidth]{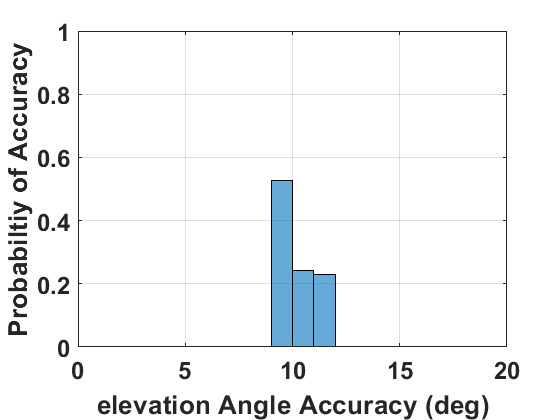}
         \caption{}
         \label{act2x2}
     \end{subfigure}
     \hfill
     \begin{subfigure}[b]{0.3\textwidth}
         \centering
         \includegraphics[width=\textwidth]{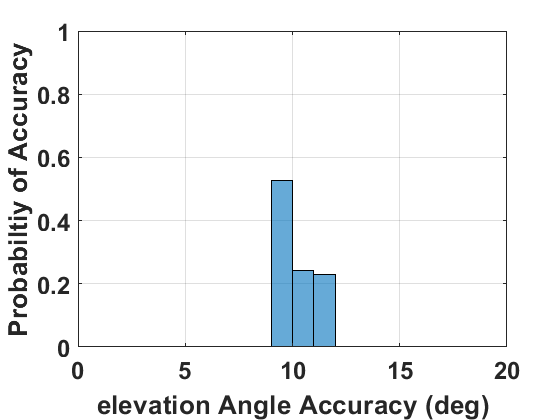}
         \caption{}
         \label{act2x8}
     \end{subfigure}
     \hfill
      \begin{subfigure}[b]{0.3\textwidth}
         \centering
         \includegraphics[width=\textwidth]{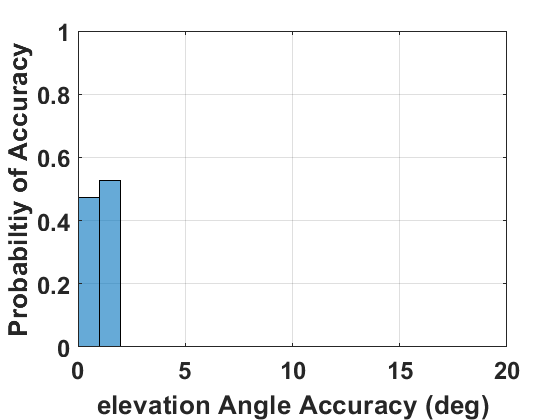}
         \caption{}
         \label{act8x8}
     \end{subfigure}
        \caption{Elevation angle estimation and accuracy for different antenna arrays using TRN-T packets (a) 2X2 estimation  (b) 2X8 estimation (c) 8X8 estimation (d) 2X2 accuracy histogram (e) 2X8 accuracy histogram  (f) 8X8 accuracy histogram.}
        \label{elt}   
\end{figure}
\\Fig.~\ref{elt} shows the elevation angle estimation using TRN-T packets. In accordance with the previous result, 8x8 antenna provides the most accurate estimation ,It also demonstrates that for similar elevation antenna angles, the results appear to be identical as of the 2x2 and 2x8 antennas. However, the 2x8 antenna provides more accurate results since it has more azimuth steering angle resulting in more directions.
\\In this chapter, we examined the impact of windowing on the parameters that were estimated and demonstrated how effective window overlap is for tracking. By using threshold sensing, we demonstrated that we can still obtain reliable estimates even when the reporting of measurements is reduced. Additionally, we demonstrated how the multi-static PPDU's performance is affected by the size of the AWV and how accurately it can estimate the angular properties of the target.

\chapter{Conclusion}
In this project, we reviewed WLAN sensing including its standardization, suggested physical layer models, sensing framework, and enabling techniques. We evaluated the sensing performance of these models using a variety of techniques, such as Doppler processing using STFT, threshold based sensing, and directional sensing. Our study showed that even with limited reported measurements, good tracking results can be achieved with overlap windowing techniques. We suggest that this work can be advanced by evaluating the multi-static PPDU's performance for collaborative sensing.

%%%%%%%%%%%%%%%%%%%%%%%%%%%%%%%%%%%%%%%%%%%

\bibliographystyle{ieeetr}

\bibliography{references}

\end{document}